\documentclass[pra,showpacs,twocolumn,nofootinbib,amsmath,amssymb,superscriptaddress]{revtex4}%
\usepackage{graphicx}

\begin{document}



\newcommand{\mub}{\ensuremath{\mu_B}}
\newcommand{\gammazr}{\ensuremath{\Gamma_{R}}}
\newcommand{\nmlfs}{\ensuremath{n_{J}}}
\newcommand{\nmhfs}{\ensuremath{n_{-J}}}
\newcommand{\DmJ}{\ensuremath{\Delta m_J}}
\newcommand{\g}{\ensuremath{\gamma}}
\newcommand{\bvec}[1]{\ensuremath{{\bf #1}}}
\newcommand{\he}{$^3$He}
\newcommand{\Rcell}{R_{\rm cell}}
\newcommand{\Lcell}{L_{\rm cell}}

\newcommand{\fig}[1]{Fig.~\ref{#1}}
\newcommand{\eqn}[1]{(\ref{#1})}

\newcommand{\mlfs}{\ensuremath{m_J=J}}
\newcommand{\mhfs}{\ensuremath{m_J=-J}}
\newcommand{\taulfs}{\ensuremath{\tau_{J}}}
\newcommand{\tauhfs}{\ensuremath{\tau_{-J}}}

\newenvironment{mytabular}[1]
{\begin{tabular}{#1} \toprule}
{\botrule \end{tabular}}

\newcommand{\stretchinclude}[1]{\includegraphics[width=3.2in]{#1}}


\title{Zeeman Relaxation of Cold Atomic Iron and Nickel in Collisions with $^3$He}

\author{Cort Johnson}
\author{Bonna Newman}%
\author{Nathan Brahms}
\affiliation{Massachusetts Institute of Technology}%
\affiliation{Harvard / MIT Center for Ultracold Atoms}
\author{John M. Doyle}
\affiliation{Harvard University}%
\affiliation{Harvard / MIT Center for Ultracold Atoms}
\author{Daniel Kleppner}
\author{Thomas J. Greytak}
\affiliation{Massachusetts Institute of Technology}
\affiliation{Harvard / MIT Center for Ultracold Atoms}
\date{\today}

\begin{abstract}
We have measured the ratio $\gamma$ of the diffusion cross-section to the angular momentum reorientation cross-section in the colliding Fe--$^3$He and Ni--$^3$He systems.  Nickel (Ni) and iron (Fe) atoms are introduced via laser ablation into a cryogenically cooled experimental cell containing cold ($< 1$ K) $^3$He buffer gas.  Elastic collisions rapidly cool the translational temperature of the ablated atoms to the \he\ temperature.  $\gamma$ is extracted by measuring the decays of the atomic Zeeman sublevels.  For our experimental conditions, thermal energy is comparable to the Zeeman splitting.  As a result, thermal excitations between Zeeman sublevels significantly impact the observed decay.  To determine $\g$ accurately, we introduce a model of Zeeman state dynamics that includes thermal excitations.  We find $\gamma_{\rm Ni-^3He} = 5\times 10^3$ and $\gamma_{\rm Fe-^3He} \leq 3\times 10^3$ at 0.75 K in a 0.8 T magnetic field.  These measurements are interpreted in the context of submerged shell suppression of spin relaxation as studied previously in transition metals and rare earth atoms. \cite{doyleTransitionMetals,doyleRareEarths,newTmTheory}.  
\end{abstract} 

\maketitle

\section{\label{sec:Intro}Introduction}

Cooling and trapping atoms at cold and ultracold temperatures has led to unprecedented control of the external and internal atomic degrees of freedom.  Such control has aided in the advance of quantum information, precision measurement and atomic clocks, and has set the stage for quantum simulation of condensed matter systems.  Successful evaporative cooling of trapped, dilute atomic gases requires a thorough understanding of elastic collisions that thermalize the sample and inelastic processes that cause heating and atom loss.  Efficient evaporation is possible only for species with a high ratio of elastic to inelastic collisions \g.  For atoms with isotropic interactions, such as the alkalis, \g\, is typically large.  Alkalis can also be conveniently laser cooled due to their simple electronic structure.  For these reasons, alkalis have been the atoms of choice for the majority of ultracold atomic physics experiments.  

Because of the proliferation of applications for ultracold atoms, there is a natural desire to develop techniques to expand the class of atomic species that can be cooled and trapped to those with rich electronic structures.  Recently, alkaline earth metal atoms and atoms with similar electron structure have been cooled to quantum degeneracy using a combination of laser cooling and evaporation out of an optical dipole trap.  Examples include $^{40}$Ca \cite{CaBEC}, $^{84}$Sr \cite{SrBECKillian,SrBECInnsbruck}, and multiple ytterbium isotopes \cite{Takahashi170YbBEC,TakahashiFermi,TakahashiMottInsulator,TakahashiYbBEC,TakahashiYbMixtures}.  There is also great interest in systems with large anisotropic magnetic dipole-dipole interactions, which have been predicted to have novel quantum phases \cite{Lewenstein:DipolePhases}, have potential use in quantum computing \cite{QCwithMagDipole}, and demonstrate geometry-dependent BEC stability \cite{DipolarGases:PRL,Pfau:dipolarStability}.  Although metastable states of alkali earth atoms are one candidate to demonstrate the above effects \cite{Alkaline-EarthTrapping}, the measured inelastic loss rates in trapped samples are high \cite{prl:yamaguchi:YB,Sr88Collisions,CaMetastableInelastic}, making evaporative cooling difficult.  Identifying viable alternatives with even larger magnetic moments requires the development of more general methods of trapping and cooling exotic species.

One technique capable of trapping exotic species is buffer-gas cooling \cite{bufferGas1}.  In buffer-gas cooling, elastic collisions with a cold buffer gas, typically helium, are used to cool the atom or molecule of interest to low enough temperatures that the species may be magnetically trapped.  Traditional evaporative cooling techniques may then be used to cool the species to the regime of ultracold temperatures.  The flexibility of buffer-gas cooling has been demonstrated by the successful trapping of many non-alkali atomic and molecular species \cite{bufferGasPrimer,doyleRareEarths,NateCuAg,MnTrapping,HeEvapBG,CindyMo,MattN,WesNH}.  Furthermore, a buffer-gas cooled Bose-Einstein condensate of metastable helium ($^4{He}^*$) was recently reported \cite{DoretBEC}, demonstrating the viability of the technique to create ultracold gases.  Buffer-gas loading of magnetic traps requires about 50 collisions in order for the target species to fully thermalize with the buffer gas.  However, it takes additional time for the atoms to diffuse toward the center of the trap and for the buffer gas to be removed.  Thus buffer-gas loading of magnetic traps has required that the orientation of the species' magnetic moment must be preserved for at least $10^4$ collisions \cite{JinhaThesis}.  

To first order, angular momentum reorientation in such collisions is dominated by the interaction between the orbital angular momentum $\bvec{L}$ of the species and the angular momentum $\bvec \ell$ of the collision \cite{romanPState}.  One therefore expects weak reorientation for species having $L=0$, and relatively strong reorientation for species having $L\neq 0$.  For instance, oxygen ($L=1$) reorients its angular momentum in almost every collision with low-temperature He \cite{romanPState}, while atoms such as potassium or copper ($L=0$) have been shown to survive more than $10^6$ collisions without angular momentum reorientation \cite{NateCuAg,NateLiK}.  Recent work \cite{romanRareEarths,romanTransitionMetals}, however, has demonstrated the existence of a class of atoms, dubbed the ``submerged shell'' atoms, in which the $L\neq 0$ valence shell is protected from angular reorientation by outer-lying filled shells of electrons.  In these species, the anisotropy of the interaction with helium is highly suppressed and $\g$ is orders of magnitude larger than typically found in non-S-state atoms.  Experimental work with rare earth atoms \cite{doyleRareEarths} and the group 3--5 transition metals \cite{doyleTransitionMetals} has shown the ratio $\gamma$ of diffusion cross-section to angular momentum reorientation cross-section to be between $10^4$ and $10^5$ in these ``submerged shell'' atoms.  Similar suppression of Zeeman relaxation has been observed in collisions between helium and $^2P_{1/2}$ species due to their spherically symmetric electron-density distribution \cite{WeinsteinDoubletPOneHalfZRSuppression}.  Suppression of fine-structure-changing collisions has also been observed in collisions between ``submerged shell" transition metal titanium with helium \cite{WeinsteinTiJChanging}.   

Previously only a few transition metals were studied \cite{doyleTransitionMetals} and each species had a small magnetic moment ($\leq 1.32$ bohr magnetons).  Because species with large magnetic moments have stronger dipole-dipole interactions and are easier to trap, characterizing transition metals with larger magnetic moments is desirable.  Our apparatus is equipped with a large cryogenic valve that can be rapidly opened to remove the buffer gas after initial cooling, allowing us to thermally isolate and evaporatively cool a trapped sample.  Thus we are well positioned to study species with $\g$ in the range of previously studied ``submerged shell'' atoms.  

Our goal was to investigate the suppression of reorientation for transition metals with large magnetic moments in their collisions with low-temperature \he.  Specifically, we study $\g$ for the Ni--$^3$He and Fe--$^3$He systems because nickel and iron have strong spectroscopic lines accessible to our doubled dye laser system\footnote{Cobalt was not included in this study as the large hyperfine structure of the atom precluded spectroscopic identification of the atom's individual Zeeman states.}.  We find that these atoms have $\gamma \lesssim 5\times 10^3$.  As with previously studied transition metals, their reorientation is more rapid than rare earth ``submerged shell'' species and is, unfortunately, too rapid for further study in our apparatus.  However, we find that nickel ($\g \sim 5 \times 10^3$) still demonstrates significant reorientation suppression compared to strongly anisotropic species.

When the thermal energy of the atomic sample is comparable to the energy separation between adjacent Zeeman states, thermal excitations have a non-negligible effect on Zeeman state dynamics.  This manifests itself as a decrease of the observed decay rate compared to the angular momentum reorientation decay rate.  As a result, a na\"\i ve model that neglects these effects will lead to an overestimation of $\gamma$.  We introduce a method for extracting $\g$ from observed Zeeman state decay by including thermal excitations in our Zeeman state dynamics model.  We also discuss the impact of our method on the interpretation of previous measurements of $\g$ for transition metals and rare earth atoms. 

\section{Experimental Methods}
\label{sec:ExperimentalMethods}

We are interested in the ratio between the atom--He diffusion cross-section $\sigma_d$ and the angular momentum reorientation cross-section $\sigma_R$.  We are specifically concerned with reorientation from the most low-field seeking trap state, $m_J = J$, to any other Zeeman state.  In the presence of a confining magnetic field, these other states will leave the magnetic trap on a time scale exponentially faster than the lifetime of the $m_J=J$ state.  We therefore write
\begin{gather}
\sigma_R \equiv \sum_{m_J' \neq J} \sigma_{J \rightarrow m_J'}\label{eq:sigma_R}, \\
\gamma \equiv \frac{\sigma_d}{\sigma_R}\label{eq:gamma}.
\end{gather}
The diffusion cross-section may be measured by observing the diffusion of the atoms to the wall when no magnetic field is present.  To measure $\sigma_R$, we apply an approximately uniform magnetic field to separate the various $m_J$ sublevels of the atom, then measure the time constant for loss of the $m_J = J$ population as a function of the zero-field diffusion time.  Extraction of the cross-section ratio from measurements of the atom decay time constants is discussed in \S\ref{sec:LifetimeModels}.

\begin{table}
\centering
\begin{mytabular}{@{\extracolsep{1ex}}cccc}
Atom & Configuration & Term & Moment ($\mu_B$) \\
\colrule
Fe & [Ar].$3d^6$.$4s^2$ & $^5D_4$ & 6.005\\
Ni & [Ar].$3d^8$.$4s^2$ & $^3F_4$ & 5.002\\
\end{mytabular}
\caption{Electronic configurations of iron and nickel.}
\label{tab:FeNi}
\end{table}


\subsection{Cryogenic Apparatus}
\label{subsec:cryoApparatus}

\begin{figure}
	\includegraphics{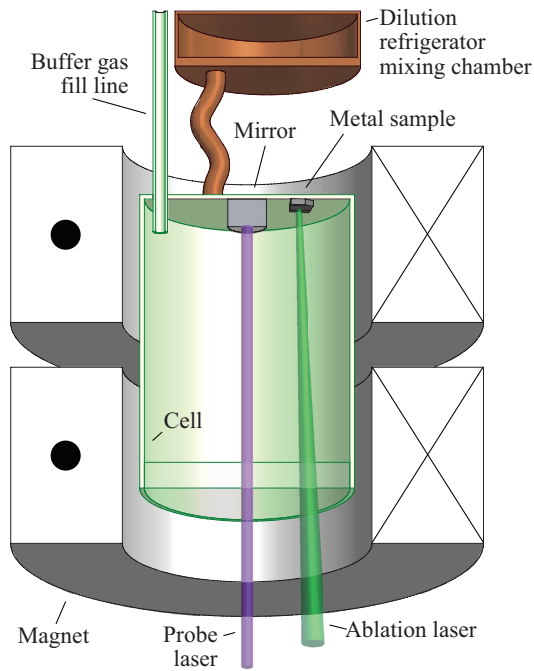}
	\caption{(color online) Schematic drawing of the cryogenic apparatus.  The magnet can operate in Helmholtz (shown) or anti-Helmholtz configurations.}
	\label{fig:schematicColor}
\end{figure}

A schematic representation of our apparatus (not drawn to scale) is shown in \fig{fig:schematicColor}.  The body of the experimental cell is machined from a G10 tube 7.6 cm in diameter and 30 cm in length.  It is thermally anchored to the mixing chamber of a dilution refrigerator by four half-inch diameter oxygen free high purity copper braids.  Thermal conductivity along the length of the cell is provided by $\sim$ 1,000 0.25 mm diameter copper wires running vertically along the outer G10 cell wall.  Each wire is electrically insulated from the rest to prevent eddy current heating as we ramp magnetic fields.  Base temperature of the refrigerator is 30 mK resulting in a cell top temperature of $\sim$ 100 mK and cell bottom temperature of $\sim$ 170 mK.  A 6.6 cm sapphire window forms the bottom of the cell.  Sapphire passes the UV frequencies needed for spectroscopy and has higher thermal conductivity than fused silica.  The window is epoxied into the G10 body, forming a vacuum seal with the cell wall.    

The cell body is housed in a cylindrical vacuum chamber with 1 mm radial clearance.  A superconducting magnet fits tightly around the vacuum chamber \cite{HarrisRSIMagnets}.  Clearances are made as tight as possible to enable maximum fields at the cell wall.  The room temperature connections to the magnet coils can be wired to produce a Helmholtz field, providing a uniform field within the cell, or an anti-Helmholtz field, providing a 4 T deep spherical quadrupole trap within the cell.  This work required only Helmholtz fields. 

Buffer gas is introduced into the cell through a fill line.  The fill line is connected to a vacuum chamber containing a charcoal sorb filled with $^3$He.  The vacuum chamber is thermally anchored to a 4K helium bath.  For each buffer-gas load, we heat the sorb with a resistive heater to $\sim$ 10 K.  This drives helium off the sorb, through the fill line, and into the cell.  Adjusting the duration and power of the heater pulse allows us to vary the amount of loaded buffer gas in a predictable manner.  Before each decay lifetime measurement is performed, the cell is heated to $\sim$ 350 mK.  This drives the buffer gas off the walls in preparation for introducing the atomic species.  

\subsection{Spectroscopy}
\label{subsec:Spectroscopy}

Fe and Ni atoms are produced via ablation of metallic targets mounted inside the cell.  Ablation is performed with a 10 ns pulse from a doubled YAG laser operating at 532 nm.  Both atom density and temperature increase with ablation power.  We used pulse energies $\sim 15$ mJ to reach densities that yielded adequate signals.  Unfortunately, these powers also resulted in temperatures at which thermal excitations between $m_J$ states significantly contributed to the observed decay rates.  

We probe atomic density, lifetime, and temperature via balanced absorption spectroscopy on the $^5D_4 \rightarrow {^5F_5}$ transition at 248 nm in Fe and the $^3F_4 \rightarrow {^3G_5}$ transition at 232 nm in Ni.  The light in both cases was produced from a dye laser doubled in a resonant cavity containing a BBO crystal.

Optical access into the cryogenic apparatus is limited to a single port through the bottom.  Beam steering optics mounted to the bottom of the dewar direct the laser into the cell.  The beam retroreflects from a mirror at the top of the cell and the exiting light is detected on a photomultipler tube (Hamamatsu H6780-04 \cite{Hamamatsu}).


\subsection{Measurement of the diffusion lifetime}
\label{subsubsec:characterizenBG}

The lifetime of the atoms at zero-field is set by their diffusion through the buffer gas to the cell wall.  The diffusion lifetime $\tau_d$ is proportional to the buffer-gas density $n_b$ in the cell (see equation \ref{eq:tauDnBG}).  A measurement of $\tau_d$ is therefore a relative measurement of $n_b$. 

To make the lifetime measurement the frequency of the laser is tuned to the atomic resonance of interest.  The laser frequency is scanned repeatedly over the absorption feature as atoms are introduced into the cell.  The number of absorbing atoms in the probe beam is measured by integrating the spectrum over the atomic line.  Drifts in the other sources of loss in the optical path can be mitigated by subtracting the signal at a dark portion of the spectrum from this integral.  We fit the decay in this integrated optical depth to a single exponential decay function to determine the diffusion lifetime $\tau_d$.  Because the laser scan has a maximum bandwidth of $\sim 30$ Hz, decays faster than this are measured by parking the laser frequency at the absorption peak.  However, this latter method is subject to low frequency noise and drift.


\subsection{Measurement of the \mlfs\ lifetime}
\label{subsubsec:measureTauMLFS}

To measure the lifetime $\taulfs$ of the \mlfs\ state, we must resolve an absorption peak from the \mlfs\ ground state.  This is accomplished by turning on a homogeneous magnetic field, thereby splitting the atom's Zeeman sublevels.  The lines are also broadened, due to field inhomogeneity within the magnet.

Consider a transition from a $m_J$ ground state with Land\'e factor $g_J$ to a $m_J + \Delta m_J$ excited state with Land\'e factor $g_J'$.  The field induced frequency shift ${\Delta}\nu_Z$ is
\begin{equation}
{\Delta}\nu_Z = \left ( g_J^{'} \Delta m_J + (g_{J}^{'}-g_{J}) m_J \right )\frac{\mu_{B}B}{h}
\label{eq:DeltanuZ} 
\end{equation}
where $h$ is Planck's constant.  Selection rules require ${\Delta}m_J = 0,\pm 1$. The first term in \eqn{eq:DeltanuZ} is generally much larger than the second, yielding three manifolds of equally spaced peaks, each corresponding to a ground state Zeeman sublevel.

\begin{figure}
\stretchinclude{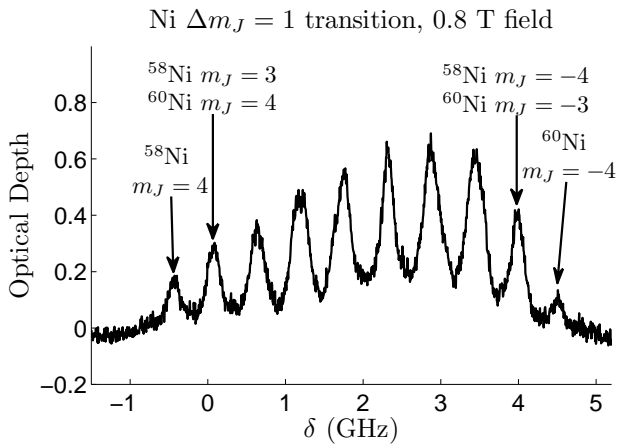}
\caption[Nickel spectrum in Helmholtz field]{Nickel optical depth vs.\ frequency (arbitrary zero) in a homogeneous (Helmholtz) field.  Each isotope has 9 lines corresponding to the 9 $m_J$ states.  The Zeeman and isotope shifts are roughly equal at 0.8 T fields, causing lines of different isotopes to overlap.  Measurements of the $m_J = J$ state lifetime are performed by parking on the $m_J = 4$ transition peak and measuring the optical depth vs.\ time.}
\label{fig:NiZeemanPeakIDExpt}
\end{figure}

The \mlfs\ state is identified by tuning to the ${\Delta}m_J = 1$ set of peaks as shown in \fig{fig:NiZeemanPeakIDExpt}.\footnote{
Clebsch-Gordan coefficients suppress the \mlfs\ peak in the $\Delta m_J=-1$ and $\Delta m_J=0$ manifolds.
}
Nickel's spectrum in a magnetic field is relatively simple because the most common isotopes, $^{58}$Ni,$^{60}$Ni, and $^{62}$Ni, have no nuclear spin and hyperfine effects are absent.  Each isotope splits into 9 lines, corresponding to the $m_J$ sublevels.  The Zeeman splitting at $B \sim$ 0.8 T is approximately equal to the isotope shift between $^{58}$Ni and $^{60}$Ni.  As a result, transitions from the $^{58}$Ni $m_J$ state overlap with $^{60}$Ni transitions from the $m_J + 1$ state.  Only the $^{58}$Ni $m_J = J$ and $^{60}$Ni $m_J = -J$ states do not experience any overlap.  We measure $\taulfs$ by tuning the laser frequency to the fully resolved $^{58}$Ni $m_J = 4$ absorption peak and observing the decay of the absorption signal.  The most abundant iron isotopes ($^{56}$Fe and $^{54}$Fe) are also $J=4$ species without hyperfine structure, so the spectroscopic methods outlined above also apply to Fe.  We choose to operate at 0.8 T with Ni and 1.0 T with Fe, as these are the highest fields at which the spectra are easily interpreted and at which the absorption lines are not too broadened to achieve a good signal-to-noise ratio.

Measuring $\taulfs$ was particularly difficult for iron because the observed lifetimes were very short at all buffer-gas densities.  Immediately after ablation we typically see an absorption signal that decays in 2-3 milliseconds.  This signal occurs at all buffer-gas densities and is present when the laser is parked near, but not necessarily on, a Zeeman level resonance peak.  We associate this transient signal with the decay of higher order diffusion modes.  Because it is unrelated to the momentum reorientation we are trying to measure, we typically ignore data taken in the first 10 ms after ablation.  For example, Ni $\taulfs$ were found using data taken between 10 ms -- 50 ms after ablation.  However, due to Fe's rapid Zeeman relaxation, the absorption signal was too small to be measured after 10 ms.  We therefore measured $\taulfs$ for Fe starting only 5 ms after ablation.

\begin{figure}
\stretchinclude{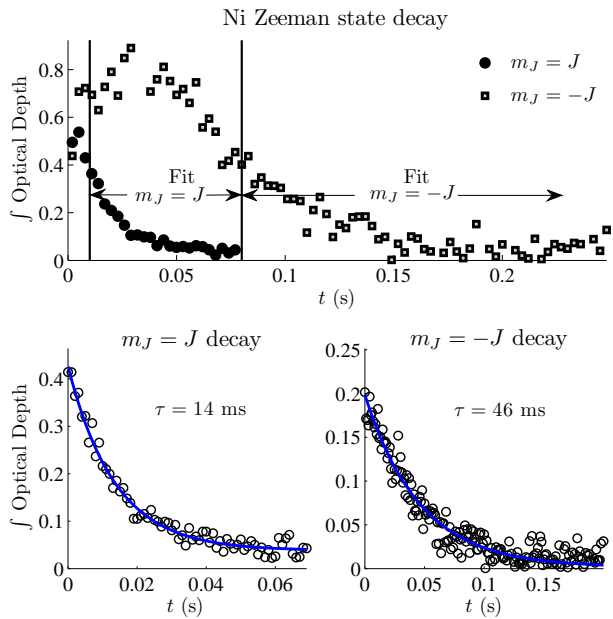}
\caption{(color online) Nickel \mlfs\ and \mhfs\ optical depth (integrated over the atomic line) vs.\ time.  Upper plot shows Zeeman state evolution starting immediately after laser ablation.  The \mhfs\ state experiences an initial increase in optical depth because $m_J \neq -J$ states are relaxing into the \mhfs\ state.  Lifetime fits are performed over the indicated regions.  Lower left(right) plot shows \mlfs($-J$) lifetime fit.}
\label{fig:NiDecayPlotsRawAndFits}
\end{figure}

Using the above procedure we can also study the dynamics of the $m_J = -J$ most high-field seeking state.  After several Zeeman relaxation times have elapsed, thermal equilibrium is established between the Zeeman and kinetic degrees of freedom and the $\mhfs$ state decays via diffusion.  We measure $\tauhfs$ at each buffer-gas density by observing $m_J = -J$ decay after thermal equilibrium has been established.  We then compare $\tauhfs$ with $\tau_d$.  Under these conditions we expect $\tauhfs$ to equal $\tau_d$.  \fig{fig:NiDecayPlotsRawAndFits} shows an example of \mlfs\ and \mhfs\ state decay with fits to a single exponential lifetime.

\subsection{Temperature Measurement}
\label{subsubsec:TempMeas}

\begin{figure}
\stretchinclude{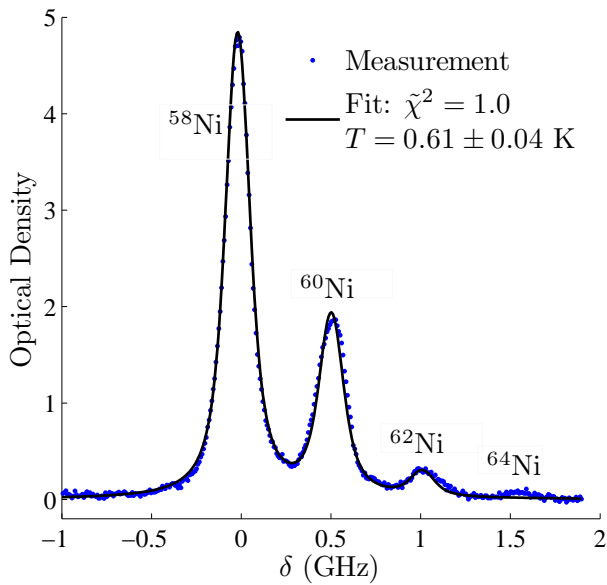}
\caption[Nickel zero-field spectrum 50 ms after ablation.]{(color online) Nickel zero-field optical depth vs.\ frequency 50 ms after ablation.  Frequency zero set to $^{58}$Ni resonance.  Temperature and density are found by fitting to a Voigt profile.}
\label{fig:NiZeroField50ms}
\end{figure}  

At atom temperatures greater than or comparable to the atomic Zeeman splitting, 670 mK at 0.8 T for Ni, thermal excitations cause the observed loss rate of the \mlfs\ state to differ from the Zeeman relaxation rate.  In order to know if these excitations can be ignored, we determine atom temperature by measuring the broadening of the zero-field spectrum.
\fig{fig:NiZeroField50ms} shows the zero-field spectrum of the $a^3F_4 \rightarrow y^3G_5$ transition of Ni at 232 nm, taken 50 ms after ablation.  The optical detuning is calibrated using a Fabry-Perot cavity.  The observed atom density is $\sim$ 3 $\times$ 10$^8$ cm$^{-3}$, corresponding to 3 $\times$ 10$^{10}$ Ni atoms in the cell.  The temperature of the atoms is determined by fitting to a Voigt profile \cite{Demtroder}.

\begin{figure}
\stretchinclude{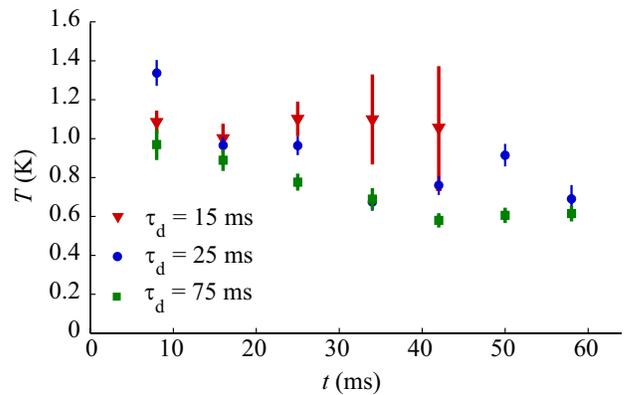}
\caption[(color online) Nickel temperature vs.\ time]{(color online) Nickel temperature vs.\ time.  Temperature measurements were made at 3 different buffer-gas densities.  The atoms cool only slightly over the time scale of our $\taulfs$ measurements.  The temperature is slightly higher at lower buffer-gas densities.}
\label{fig:NiTempFitsVsTime}
\end{figure}

\fig{fig:NiTempFitsVsTime} shows the temperature of the Ni atoms as a function of time for three buffer-gas densities.  Thermal excitations have the greatest impact at high buffer-gas densities, where the Zeeman relaxation rate is much faster than the diffusion rate.  In our analysis, we shall use the average measured Ni temperature at high densities, 750 mK.

The zero-field spectrum could potentially be broadened due to trapped fluxes in the superconducting magnet.  These have previously been measured to be $\leq 10$ gauss \cite{OurMagnet}, yielding a systematic uncertainty of $_{-110}^{+0}$ mK.

\section{Lifetime Models}\label{sec:LifetimeModels}

 
\subsection{Model in the near-zero temperature limit}
\label{subsec:MLFSModels}

At high buffer-gas densities and in the limit of near-zero temperature, atoms are lost from the \mlfs\ state due to two primary mechanisms.  First, atoms diffuse via elastic collisions until they reach the cell walls, where they stick.  Second, atoms experience Zeeman relaxation.  In this section we ignore collisions that repopulate the \mlfs\ state once the atom has relaxed into a lower energy state.  We will include this finite temperature effect in Section \ref{subsec:ZeemanCascade}.

In a cylindrical cell of radius $\Rcell$ and length $\Lcell$, the lifetime due to diffusion for atoms in the lowest order diffusion mode is \cite{hastedCollisions}
\begin{gather}
\tau_d =  \frac{n_{b}\,\sigma_d}{\bar{v}\,{G}},
\label{eq:tauDnBG} \\
G = \frac{3\pi}{32}\, \left ( \frac{\pi^2}{\Lcell^2}+\frac{j_{01}^2}{\Rcell^2} \right )
\end{gather}
\noindent where $j_{01}=2.40483\ldots$ is the first zero of the Bessel function $J_0(z)$, $n_b$ is the buffer-gas density, and $\bar v = {({8k_BT}/{\mu\pi})^{1/2}}$ is the mean relative velocity of the colliding atom--$^3$He system with reduced mass $\mu$.  We can ensure that all atoms are in the lowest order diffusion mode by a waiting for a few multiples of $\tau_d$ before measuring the atom lifetime.

Near zero temperature, the lifetime of atoms in the \mlfs\ state is the reciprocal sum of the diffusion lifetime and the lifetime due to Zeeman relaxation:
\begin{equation}
	\taulfs = \left ( \frac{1}{\tau_d} + \frac{1}{\bar v\,\sigma_R\,n_b} \right )^{-1}.
	\label{eq:tauLfsNbg}
\end{equation}
Unfortunately we do not have an absolute calibration of $n_b$.  However, we can use \eqn{eq:tauDnBG} to eliminate $n_b$ from \eqn{eq:tauLfsNbg}.  Doing this, and replacing $\sigma_d/\sigma_R$ with $\gamma$, we find
\begin{equation}
	\taulfs = \frac{\tau_d}{1+ \tau_d^2 \, \bar v^2 \, G / \gamma}.
	\label{eq:tauLfsZeroT}
\end{equation}
$\gamma$ can be found by measuring $\taulfs$ as a function of $\tau_d$ (measured at zero magnetic field), and fitting to \eqn{eq:tauLfsZeroT}.  However, as we shall show in the next section, this model breaks down when $k_B T\gtrsim g_J \mu_B B$.


\subsection{Model at finite temperature}
\label{subsec:ZeemanCascade}

The previous model assumes that once a \mlfs\ atom experiences Zeeman relaxation it remains in a lower energy Zeeman state forever; the possibility of excitation \emph{into} the \mlfs\ states was ignored.  When the thermal energy $k_BT$ is much less than the magnetic interaction energy, this assumption is valid as collisions do not have sufficient energy to excite atoms into states with higher $m_J$.  However, for our experimental conditions $k_B T << g_J\mu_{B}B$ does not hold, and a non-negligible percentage of collisions with the buffer gas have enough energy to excite an atom into a higher energy Zeeman state.  This effect slows the observed loss from the \mlfs\ state, yielding an overestimate of $\gamma$ when the near-zero temperature model is used.  Therefore, a correct determination of $\gamma$ from measured data must take thermal excitations into account.  

For the above reasons a Zeeman cascade model which includes diffusion, the dynamics of all Zeeman states, and the possibility of thermal excitation must be developed.
In our model, the density $n_{m_J}$ of each $m_J$ level with energy $E_{m_J}$ at temperature $T$ evolves according to
\begin{equation}
\label{eq:FullDynamics}
\begin{split}
\dot{n}_{m_J} &= -\Gamma_{d}\,n_{m_J}-\Gamma_{R}{\sum_{{m_J'} < {m_J}} \alpha_{{m_J' m_J}} \, n_{m_J}}\\
& -\Gamma_{R}{\sum_{{m_J'} >{m_J}}\alpha_{m_J' m_J} \, n_{m_J}}
\exp\left(-\frac{(E_{{m_J'}}-E_{m_J})}{k_{B}T}\right)\\
& + \Gamma_{R}{\sum_{{m_J'} > {m_J}} \alpha_{m_J' m_J} \, n_{{m_J'}}}\\
& + \Gamma_{R}\sum_{{m_J'} < {m_J}}\alpha_{m_J' m_J} \, n_{{m_J'}}\exp\left({-\frac{(E_{m_J}-E_{{m_J'}})}{k_{B}T}}\right).
\end{split}
\end{equation}
where $\Gamma_d \equiv 1/\tau_d$ and $\Gamma_R \equiv 1/\tau_R \equiv \sigma_R n_b \bar v$ are the diffusion and relaxation rates.  $\gamma$ is related to these quantities by $\gamma = \tau_R \, \tau_d \, G \, \bar{v}^2$.  $\alpha_{m_J' m_J}$ in each summation represents the coupling between $m_J$ and ${m_J'}$ Zeeman levels, subject to
\begin{equation}
\sum_{{m_J'} = 1}^{2J}\alpha_{m_J' m_J} = 1,
\end{equation}
and
\begin{equation}
\alpha_{m_J' m_J} = \alpha_{m_J m_J'}
\end{equation}
The first term in \eqn{eq:FullDynamics} is diffusion loss, the second is Zeeman relaxation into lower energy states, the third is thermal excitation into higher energy states, the fourth is Zeeman relaxation from higher energy states, and the fifth is thermal excitation from lower energy states.

The dominant effect of thermal excitation on \mlfs\ atoms is an alteration of their initial decay from the simple exponential predicted by the near-zero temperature model.  At early times ($t\ll 1/\Gamma_R$, when all $m_J$ states are equally populated), thermal excitations cause the \mlfs\ state to decay more slowly than the prediction of \eqn{eq:tauLfsZeroT}.  The exact scaling will depend on $T$, $B$, and the exact form of $\alpha_{m_J' m_J}$.  At later times ($t>(2J+1)/\Gamma_R$), the Zeeman states approach thermal equilibrium, and the lifetimes of all $m_J$ levels approach $\tau_d$.

Finding reliable values of $\alpha_{m_J' m_J}$ presents a challenge \cite{romanPersonal}.  Because the rates for iron and nickel are unknown, they must be estimated.  The assumptions adopted significantly affect the predicted Zeeman state dynamics.  Consequently, we have analyzed them using three scenarios for transitions between Zeeman states: all transitions between Zeeman states are equally allowed, only ${\Delta}m_J = \pm 1$ transitions are allowed, and an intermediate regime based on calculations for thulium and general $^3P$ atoms \cite{romanTmHe,romanPState}.  The relative values of $\alpha$ used in this third case are shown in Table~\ref{tab:SelectionRules}.

\begin{table}
\centering
\begin{mytabular}{c@{\hspace{2ex}}r@{.}l}
$\Delta m_J$ & \multicolumn{2}{l}{$\displaystyle{\frac {\alpha_{\Delta m_J}}{\alpha_{\pm 1}}} $} \\
\colrule
$\pm 1,\,2$ & \hspace{1ex}1&0     \\
$\pm 3,\,4$ & 0&2   \\
$\pm 5,\,6$ & 0&04  \\
$\pm 7,\,8$ & 0&008 \\
\end{mytabular}
\caption{Relative rate coefficients for Zeeman relaxation with a given $\Delta m_J$, based on \cite{romanTmHe,romanPState}.}
\label{tab:SelectionRules}
\end{table}


\section{Determination of $\gamma$}
\label{sec:DetermineGamma}

We use the finite temperature model to fit for $\gamma$ using the following method:  For each experimentally observed value of $\tau_d$, we simulate the \mlfs\ state decay using a guess value of $\gamma$, a temperature of 750 mK, a magnetic field of 0.8T, and the literature-based $\alpha$ values from Table~\ref{tab:SelectionRules}.  We then fit the simulated decay of the \mlfs\ state over the same time interval used to measure $\taulfs$.  Finally, we perform a $\chi^2$ fit of the simulated values of $\taulfs$ to measured values to determine $\gamma$.


\subsection{$\gamma$ for Ni--\he\ collisions}

\begin{figure}
\stretchinclude{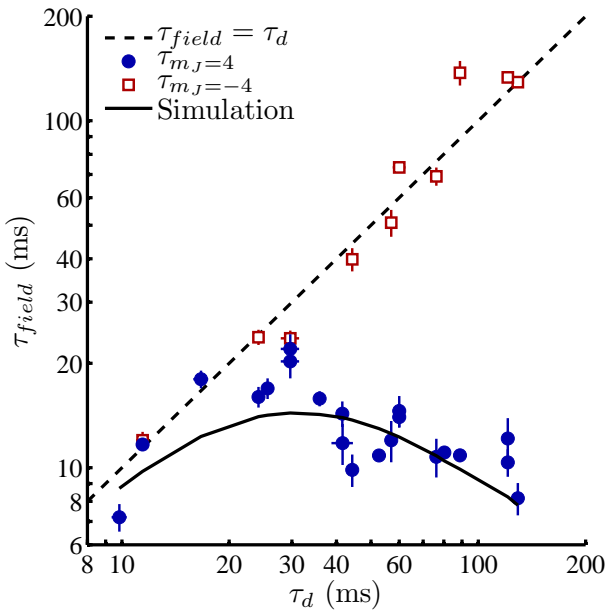}
\caption[Nickel data with fit to Zeeman cascade simulation.]{(color online) Nickel \mlfs\ and \mhfs\ lifetimes in a 0.8 Tesla field vs.\ diffusion lifetime.  The dashed line was drawn with a slope = 1 to demonstrate that the \mhfs\ atoms leave the cell by diffusion as expected.  The \mlfs\ atoms decay quickly due Zeeman relaxation in the region of high $\tau_d$.  The best fit of the \mlfs\ data to a finite temperature Zeeman cascade simulation yields a value of $\gamma = 5 \times 10^3$.}
\label{fig:NiFinalResult}
\end{figure}

The \mlfs\ state lifetimes are plotted with the \mhfs\ state lifetimes in \fig{fig:NiFinalResult}.  As expected, the \mhfs\ state lifetimes are approximately $\tau_d$, whereas the \mlfs\ state lifetimes first increase, then decrease with increasing $\tau_d$.  The clear differentiation between the \mlfs\ state and \mhfs\ state behaviors provides convincing evidence that we are measuring Zeeman relaxation of the \mlfs\ state.

\fig{fig:NiFinalResult} shows the best fit of the \mlfs\ data to the finite-temperature model, yielding $\g = 5 \times 10^3$.  This is nearly a factor of two smaller than the $\g = 9 \times 10^3$ obtained when fitting to the near-zero temperature model.  Uncertainty in $\g$ results from three primary effects: deviation of the data from the model, uncertainty in thermal excitation rates due to temperature uncertainty, and uncertainty in the assumed ``selection rules" for Zeeman relaxation.

Our data generally show deviations from the model larger than their statistical uncertainties.  At low $\tau_d$, we believe this is due to an unknown experimental artifact, as was also observed in our experiments with Cu--He and Ag--He \cite{NateCuAg}.  At high $\tau_d$, deviation is caused by performing measurements in a regime where loss is not purely exponential, as we begin to see the atomic states approach their steady-state distributions.  We account for the effect of this model deviation on our measurement of $\g$ using an F test \cite{Bevington}, yielding a parameter uncertainty with a 95 \% confidence interval of $^{+2.2}_{-1.6} \times 10^3$.

\begin{table}
\centering
\begin{mytabular}{@{\extracolsep{1ex}}cc}
Selection Rule & Fit for $\gamma$\\
\colrule
From Literature      & 5 $\times$ 10$^3$\\
All Equal            & 7 $\times$ 10$^3$\\
$\Delta m_J =1$ only & 2 $\times$ 10$^3$\\
\end{mytabular}
\caption{Values of $\gamma$ extracted from fits of data to Zeeman cascade simulations under various selection rules assumptions.  For these fits, we used $T=0.75$~K.}
\label{tab:FindGammaBounds}
\end{table}

Finally we consider the effects of the assumed selection rules for Zeeman relaxation.  For the fit in \fig{fig:NiFinalResult} we used guidance from the literature in setting realistic selection rules.  Since it is impossible to quantify the accuracy of this assumption, we systematically vary the selection rules and observe the resulting fit for $\g$.  One extreme assumption is $\Delta m_J = \pm 1$.  The opposite extreme is that the rates into all energetically allowed states are equal.  By fitting the entire data set to the simulation using the extremes in selection rules, we obtain the result in Table \ref{tab:FindGammaBounds}.  The values of $\g$ vary between $2 \times 10^3$ and $7 \times 10^3$.  We assign a lower bound of $\g > 2 \times 10^3$ based on this result.


\subsection{Iron: Upper Limit on $\gamma$}

\begin{figure}
\centering
\stretchinclude{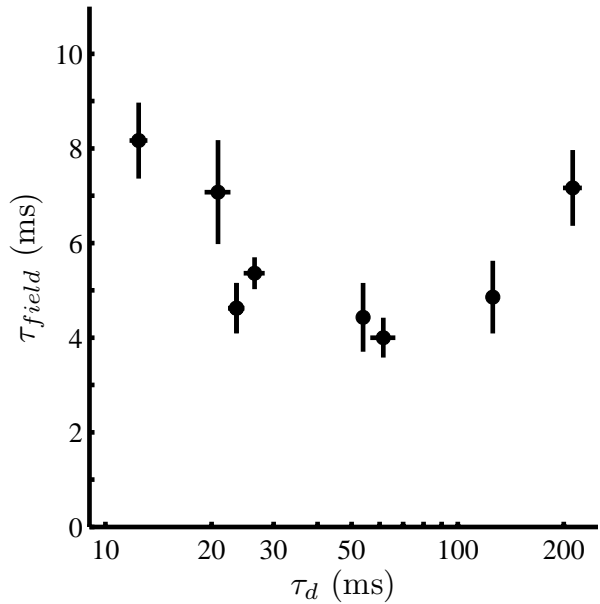}
\caption{Iron \mlfs\ lifetimes in a 1.0 Tesla field vs.\ diffusion lifetime at zero field.  There is no region of $\tau_d$ for which $\taulfs$ increases.  This indicates that Zeeman relaxation is the dominant loss mechanism for the entire range of data.}
\label{fig:FeTauMLFSvsTauD}
\end{figure}

\fig{fig:FeTauMLFSvsTauD} shows the measured $\taulfs$ vs.\ $\tau_d$.  The predicted region of $\taulfs \propto \tau_d$ at low buffer-gas density is not observed, indicating that Zeeman relaxation occurs on a time scale faster than diffusion for the entire range of $\tau_d$.  Therefore our finite temperature model cannot be used to fit for $\g$.  The slight rise in $\taulfs$ with increasing buffer gas is consistent with measuring a combination of Zeeman relaxation and diffusion as discussed in the previous section.

We set an upper bound of $\gamma < 3 \times 10^3$ by using conservative values for the parameters in the following expression:
\begin{equation} 
\gamma \leq \tau_d\tau_{R}{\overline{v}}^2{G}.
\label{eq:gammaLimit}
\end{equation}
We calculate $\overline{v}$ from the temperature measured via the Voigt profile of a zero-field spectrum taken 8 ms after ablation.  Recall that $\taulfs$ is a reciprocal sum of diffusion and Zeeman relaxation lifetimes.  When $\tau_d = \tau_{R}$ the expected value of $\taulfs = \tau_d/2$.  We assume this condition is met at our lowest buffer-gas point since $\tau_d/2$ is within the error bar of $\taulfs$.  This yields the most conservative upper bound on $\g$ because $\tau_{R}$ is at a maximum value consistent with our observation that $\taulfs < \tau_d$.  By using the above method, we are able to set an upper bound for iron that is approximately a factor of 5 smaller than the bound measured for scandium \cite{doyleTransitionMetals}.

\subsection{Impact on Previous Work}

We have demonstrated that when the Zeeman splitting between neighboring $m_J$ states is comparable to thermal energy, thermal excitations impact the value of $\g$ extracted from measurements of $\mlfs$ decay.  First, Zeeman relaxation occurs faster than the measured $\mlfs$ decay, so the extracted $\g$ is lower than it would be assuming no thermal excitations.  Second, uncertainty in selection rules for relaxation collisions leads to uncertainty in how much $\g$ must be lowered.  If we define $\xi = \mu_B g_J B/k_B T$, these effects are significant when $\xi \lesssim 1$.  The nickel measurement was taken at $\xi \sim 0.87$.  Previous work with transition metals \cite{doyleTransitionMetals} and rare earth atoms \cite{doyleRareEarths} were performed under similar experimental conditions.  We now discuss the impact of the present work on those measurements. 

The reported value of $\g$ for titanium, a $J=2$ transition metal, was found by measuring the decay of the $m_J = 2$ state.  It was assumed that the measured exponential decay rate of $\mlfs$ atoms was equal to the Zeeman relaxation rate so long as the Zeeman degrees of freedom were not in thermal equilibrium with the translational temperature $T_{\textrm{trans}}$.  Specifically, the Zeeman temperature $T_Z$ was defined at a particular field $B$ by equating the ratio of the local populations of two levels, $m_J$ and $m_J'$, to the Boltzmann factor at temperature $T_Z$:

\begin{equation}
\frac{N_{m_J}}{N_{m_J'}} = \exp\left(\frac{g_J \mu_B (m_J' - m_J) B}{k_{B}T_Z}\right),
\end{equation}

It was claimed that when $T_Z >> T_{\textrm{trans}}$, the observed $\mlfs$ decay was equal to the Zeeman relaxation rate.  Although this is a necessary condition to measure Zeeman relaxation, it is not sufficient to avoid thermal effects outlined in the present work.  The titanium measurement was performed at 3.8 T with 1.8 K atoms.  This results in $\xi = 0.94$; very similar to our measurement conditions for nickel.  Therefore, for a given set of selection rules, we expect a comparable downward shift in $\g$ which would result in better agreement between experiment \cite{doyleTransitionMetals} and theory \cite{romanTransitionMetals}.  As the number of Zeeman levels decreases, the number of decay channels decreases and the uncertainty in selection rules diminishes.  Because titanium is a $J = 2$ atom, it has five Zeeman levels compared to nine for nickel.  Therefore, the uncertainty introduced by selection rules for titanium will be less than those found for nickel.

The rare earth $\g$ values were found by measuring the decay of a trapped atomic sample.  Because multiple low-field seeking Zeeman states were simultaneously trapped in an inhomogeneous magnetic field, it was impossible to monitor the decay of the $\mlfs$ state via an isolated spectroscopic line.  As a result, a model which included thermal excitations, atom drift due to the trapping field, and diffusion was implemented to simulate the dynamics of all Zeeman states.  Therefore, the reported $\g$ values already take into account thermal effects.  However, uncertainties in selection rules were not addressed.  It was assumed that $\mlfs$ atoms could decay into any energetically allowed state with equal probability (the second row of Table \ref{tab:FindGammaBounds}).  The values of $J$ for the rare earth elements studied range from $7/2$ (thulium) to $8$ (dysprosium) compared to $4$ for nickel.  As a result, uncertainties in $\g$ due to unknown selection rules should be comparable to or worse than those found for nickel.

\section{Summary and Conclusions}

A study of collisions between the most low-field seeking Zeeman state of nickel and iron with $^3$He has been performed to determine the feasibility of buffer-gas loading highly magnetic transition metals into a magnetic trap.  Atoms were introduced via laser ablation into a cryogenic cell containing a background gas of $^3$He.  Although we could not measure buffer-gas density, its relative density was measured by observing the diffusion rate of the atomic sample through the buffer gas.  We measured the ratio $\gamma$ of diffusion cross-section to angular momentum reorientation cross-section by measuring the $\mlfs$ state decay at several different buffer-gas densities.  For our experimental conditions the energy splitting between adjacent Zeeman levels was comparable to the thermal energy of the atomic sample.  Our operating temperature was set by the ablation power required to achieve an adequate signal-to-noise ratio.  We could not operate at higher fields because magnetic broadening of the atomic resonances decreased our signal-to-noise to intolerable levels.  Under these circumstances, thermal excitations into the $\mlfs$ state cause its decay to differ from pure Zeeman relaxation.  In order to find an accurate value of $\g$, we fit measured $\mlfs$ decay to a model of Zeeman state dynamics that includes thermal excitations.  

For the Ni-$^3$He system $\gamma$ was found to be $5^{+2.2}_{-1.6} \textstyle{(stat)} \pm 1 \textstyle{(sys)} \times 10^3$, assuming the Zeeman state relaxation coupling coefficients of Table \ref{tab:SelectionRules}.  A change in the assumed relaxation coupling coefficients changes the predicted thermal excitation rates into the $\mlfs$ state, resulting in further uncertainty in $\g$ as listed in Table \ref{tab:FindGammaBounds}.  We have also set an upper limit on $\gamma$ for the Fe-$^3$He system of $3\times 10^3$.  These values of $\gamma$ are high enough to allow buffer-gas cooling to thermalize Ni and Fe, but too small to allow a sizeable sample to remain trapped after the 100 ms required for removal of the buffer gas.

Our $\g$ measurements extend the experimentally explored range of transition metal-helium collisions to species with high magnetic moments.  Our method of finding $\g$ by measuring $\mlfs$ decay at many buffer-gas densities allows us to measure smaller values of $\g$ and set smaller upper bounds than in previous work.  While we do find that inelastic angular momentum changing collisions are suppressed in the Ni-He and Fe-He systems, the degree of suppression is low compared to the rare earth elements, which have $\gamma \gtrsim 10^5$ \cite{doyleRareEarths}.  The values for $\gamma$ measured here are similar to the values measured for other transition metals, specifically the Sc-He and Ti-He systems \cite{doyleTransitionMetals}.  Our observations, together with these previous measurements, are consistent with a hypothesis of reduced collisional angular momentum transfer due to screening of the valence electrons by closed electron shells.

This work was supported by the Office of Naval Research, the National Science Foundation, and the NSF Harvard/MIT Center for Ultracold Atoms.


\end{document}